\documentclass[conference]{IEEEtran}
\IEEEoverridecommandlockouts
\usepackage{cite}
\usepackage{amsmath,amssymb,amsfonts}
\usepackage{comment}
\usepackage{algorithmic}
\usepackage{graphicx}
\usepackage{textcomp}
\usepackage{xcolor}
\usepackage{svg}
\usepackage{tabularx}
\usepackage{array}
\usepackage{subcaption}
\usepackage{xcolor}

\def\BibTeX{{\rm B\kern-.05em{\sc i\kern-.025em b}\kern-.08em
    T\kern-.1667em\lower.7ex\hbox{E}\kern-.125emX}}

\usepackage{booktabs}

\usepackage{multirow} 
\usepackage{makecell}
\usepackage{comment}
\usepackage[sort&compress,numbers]{natbib}

\begin{document}

\title{WirelessSenseLLM: Zero‑Shot Human Activity Understanding by Bridging Wireless Signals and Human Language}


\author{
\IEEEauthorblockN{Mahmuda Keya$^{1}$, Sneh Pillai$^{1}$, Jiawei Yuan$^{1}$, Kai Zeng$^{2}$, Long Jiao$^{1}$\thanks{Accepted at IEEE SECON 2026.}}

\IEEEauthorblockA{$^{1}$University of Massachusetts Dartmouth, MA, USA}
\IEEEauthorblockA{$^{2}$George Mason University, VA, USA}

\IEEEauthorblockA{\{mkeya, spillai, jyuan, ljiao\}@umassd.edu, kzeng2@gmu.edu}
}

\maketitle

\begin{abstract}
There is a growing interest in enabling wireless sensing systems to interpret human motion from unsegmented wireless signals; however, existing CSI-based applications rely heavily on accurate signal segmentation and predefined action labels, which limit their applicability in zero-shot scenarios. We present \textbf{WirelessSenseLLM}, a language-driven framework that leverages large language models (LLMs)to enable zero‐shot human motion understanding from unsegmented Wi-Fi Channel State Information (CSI). To bridge the modality gap between time-series CSI and discrete language representations, we introduce a CSI-to-Language Adapter and a cross-modal projection mechanism that allows the CSI feature to be mapped into a language-aligned semantic space. This design enables the generation of fine-grained natural language descriptions of sequential and overlapping human motions, supporting downstream reasoning without segmented training data. We address two core technical challenges: modality mismatch between CSI features and language embeddings, and overlapping actions in unsegmented CSI streams. Extensive experiments demonstrate strong performance in zero-shot action understanding (92\% accuracy and 91\% F1-score), language-based reasoning quality (30\% factual and 15\% reasoning improvements), and multi-person motion explanation with an average 12.33\% improvement over prior methods. These results highlight \textbf{WirelessSenseLLM}’s effectiveness for robust, interpretable human motion understanding from CSI signals.

\end{abstract}


\section{Introduction}

In wireless communication systems, signals propagate through the environment and undergo multi-path effects such as reflection, diffraction, and scattering. When there is any human motion, these propagation paths are continuously altered by body movements, resulting in observable variations in the received signals \cite{schafer2021human, zhao2019r}, which inherently encode information related to human motion \cite{gao2021towards}. For instance, amplitude variations in signal power are caused by interference as the human body interacts with the wireless channel \cite{gao2021towards} while phase variations reflect movement-induced fluctuations associated with spatial characteristics, e.g., direction and velocity \cite{fu2020sensing}. Due to these signal variations, wireless sensing, particularly Channel State Information (CSI), has been widely applied to human sensing tasks such as activity recognition, motion detection, and pose estimation \cite{liu2019wireless}.

To detect human actions from wireless signals, existing schemes \cite{liu2019wireless} usually follow a structured processing pipeline. It starts with capturing the raw wireless signals using commodity devices, such as WiFI routers, or specialized hardware (e.g., USRP\cite{liu2019wireless}), followed by signal processing to mitigate noise, carrier frequency error, or phase offsets. Building on these pre-processed signals, action recognition is then enabled through several additional steps. First, to establish one-on-one mapping between wireless signals and categorized human actions, sequential wireless signals \cite{bocus2022operanet} have to be accurately segmented so that each segment contains a single action. Second, segmented wireless signals will be annotated later by using the labels of different human actions \cite{liu2019wireless}. Third, to isolate the motion-related patterns, feature processing algorithms, such as FFT or wavelet transform, can be applied to extract variations that correlate with distinct human activities. 

Requiring the segmented CSI data significantly limits the applicability in real-world scenarios. When wireless signals contain continuous, unsegmented, and unlabeled streams of human motion, schemes relying on the accurate feature labeling struggle to generalize, particularly in a zero-shot scenario where unseen actions appear at test time \cite{zhang2025wi, kiani2024survey}. To fill this gap, a pre-trained LLM (e.g., GPT4o)-based scheme named \textbf{Wi-Chat} \cite{zhang2025wi} was proposed recently. \textbf{Wi-Chat} visualizes wireless signals first and then encodes information such as motion and signal behavior into prompts. Thanks to the image pattern recognition capability of GPT4o, \textbf{Wi-Chat} has an impressive performance on zero-shot scenarios (nearly 90\% of accuracy). However, \textbf{Wi-Chat} also needs accurate segmentation on wireless signals, and its performance will degrade when multiple actions are overlapped.

To address the above challenges, schemes that don't rely on accurate segmentation and annotations of wireless sensing signals are highly desired. In contrast, the stream of human motions could be decomposed into combinations of basic textual description rather than discrete labels: (a) fine-grained movement of key body parts, which includes upper body parts, lower body parts, and the torso/overall pose; (b) primitive movements (turning, stepping forward/backward, and reaching); (c) sequential compound human movements (e.g. sit down, reach out, and grab a cup) \cite{chen2024motionllm}. For example, a description such as "walking while holding an object against the ear", one can quickly infer that this person is using a phone while walking. Even if we’ve never seen this person walk, our language‐domain knowledge of human actions enables this reasoning capability. Large Language Models (LLMs) excel at such compositional, text-based reasoning. If we can teach LLM to distinguish/recognize the basic movement of key body parts and primitive movements, LLM can use cross-domain knowledge (knowledge learned by natural language) to infer higher-level actions or even compound unseen human actions. 

Inspired by this observation, the goal of this work is to develop an LLM-based wireless human sensing system that converts continuous CSI streams into sequential textual descriptions and interprets through language-based reasoning rather than predefined action labels. Specifically, the system recognizes the basic key body parts movements and primitive movements (e.g., generate the detailed textual description) for zero-shot human actions, and then utilizes language-domain knowledge in LLM to classify actions (Push-ups or squats?) or answer questions related to current human actions (How many push-ups did this person do?). In this work, we define zero-shot as the ability to understand and reason about human actions without explicit segmentation or training labels. By processing unsegmented CSI streams and generating language descriptions, LLMs can provide natural breakpoints for dividing continuous human motion streams, contextual understanding, and higher-level reasoning over human motion sequences. Thanks to the self-attention mechanism \cite{vaswani2017attention}, LLMs allow each input token to attend to every other token in the sequence regardless of distance. As a result, LLMs have the potential to establish temporal relationships of an unsegmented wireless signal stream across short and long time spans (not just frame-to-frame).

However, bridging wireless sensing signals and human language in LLMs introduces a fundamental modality mismatch challenge. Human-motion information in CSI is continuous and encodes information like Doppler shifts, signal phase, amplitude, and multi-path propagation. In contrast, human language is modeled as a discrete and symbolic token with a structured grammar. Directly mapping low-level CSI features to high-level language concepts is thus non-trivial, especially when motion patterns evolve continuously without explicit action segmentation.

To address these challenges, we propose WirelessSenseLLM, a wireless-sensing scheme built upon multimodal LLMs that accepts wireless sensing signals as input, and generates descriptions on sequential human motions and answers questions related to the human movements (i.e., a question like "what kind of actions this person is doing?"). Thanks to the above-mentioned capabilities of LLM, WirelessSenseLLM is capable of supporting zero-shot human motion understanding and in-depth reasoning. The framework employs a specialized wireless encoder to process the complex-valued CSI into sequential embeddings, a CS-to-Language Adapter to resolve the modality mismatch, and a cross-domain projection mechanism to align wireless representation into language tokens. A pretrained LLM {Vicuna-7B v1.5} is then fine-tuned to learn temporal reasoning and zero-shot inference over unsegmented CSI streams. While our approach supports both single-person and multi-person scenarios, we explicitly analyze performance degradation as interaction density increases. 


In summary, the main contributions of this work are as follows:

\begin{itemize}
    \item We introduce \textbf{WirelessSenseLLM}, a segmentation-free, language-driven framework for CSI-based human motion understanding that enables zero-shot recognition and reasoning from continuous CSI signals.  
    

    \item We propose a CSI-to-Language Adapter that bridges the gap between time-series CSI data and human language tokens, addressing the modality mismatch. 
    

    \item We introduce a new multi-person wireless sensing dataset with carefully curated natural language annotations, enabling language-driven human motion understanding from CSI data. 
\end{itemize}

\section{Related Work}\label{AA}

To enable effective wireless human motion understanding, prior work has explored non-invasive sensing modalities such as Wifi and Radar by measuring physical layer properties, including received signal strength Indicator (RSSI) and channel state information (CSI) \cite{liu2019wireless,yang2022deep}. These signals are readily available from commodity network interface cards, such as Atheors 9580 NIC \cite{xie2015precise}) or software defined radios using (FMCW) \cite{liu2019wireless}. These have been widely modeled using deep learning architectures such as CNNs \cite{zhang2020device} and LSTMs \cite{ahmad2024wifi} to capture high-dimensional temporal patterns, achieving strong performance in supervised activity recognition \cite{islam2022stc}. Unfortunately, these approaches predominantly rely on precise temporal segmentation and action-specific labels \cite{zhang2025wi}, which limits their applicability to continuous, unsegmented CSI streams and hinders zero-shot generalization. In parallel, multi-modal LLMs have demonstrated strong zero-shot reasoning in the vision domain by mapping visual data to language \cite{chen2024motionllm,lin2023video}. However, these approaches fundamentally rely on rich visual inputs and cannot be directly extended to wireless sensing. Recent work has begun to explore language-driven reasoning for wireless modalities using millimeter-wave (mmWave) radar point clouds, as demonstrated by RadarLLM \cite{lai2025radarllm}. This offers a structured spatial representation but differs significantly from CSI. Compared to radar, CSI offers broader spatial coverage, lower deployment cost, and stronger sensitivity to full-body motion \cite{wang2022skeleton}, yet its continuous, high-dimensional, and multi-path entanglement makes semantic grounding and language-based reasoning more challenging. While Wi-Chat \cite{zhang2025wi} presents an early attempt to apply LLMs to WiFi CSI through visualization and prompt-based reasoning, it still relies on accurate signal segmentation and struggles with continuous, compound motions and zero-shot reasoning from continuous CSI signals. 

Extending wireless human motion understanding to multi-person scenarios introduces additional challenges due to signal superposition and interference. Prior work addresses this setting using signal decomposition methods such as ICA to separate mixed CSI signals into independent components \cite{abuhoureyah2024multi}. Other studies employ Transformer-based architectures, often combined with CNNs or LSTMs to process long-range dependencies and separate continuous movements through self-attention mechanisms \cite{kobir2025enhancing}. While these methods show effectiveness for multi-person activity classification, they remain restricted to label-based recognition within the signal domain. Notably, although transformers are widely used as discriminative models for time series analysis, no existing work utilizes LLMs to generate, reason, or semantically interpret multi-person human motion directly from continuous CSI.

\section{Sensing Model and Problem Definition}\label{AA}
\label{sec:three}

\subsection{\textbf{Sensing Model}}

\begin{figure}[h!]
  \begin{center}
\includegraphics[width=0.48\textwidth]{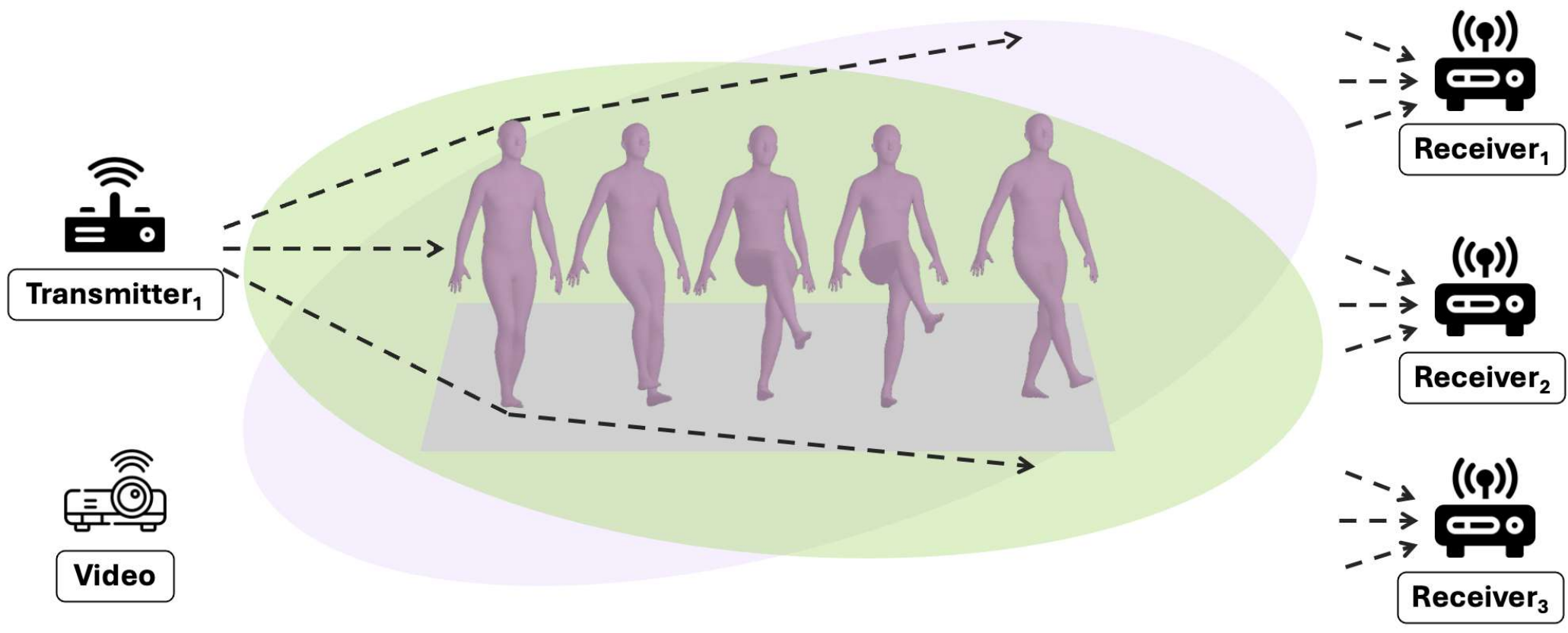}
    \caption{Illustration of WirelessSenseLLM model: one transmitter, three receivers with three antennas each to capture sequential human motion.}
    \label{system_model}
    \end{center}
\end{figure}

\begin{figure}[h!]
  \centering
  \begin{subfigure}[b]{0.49\linewidth}
    \centering
    \includegraphics[width=\linewidth,height=1.6cm]{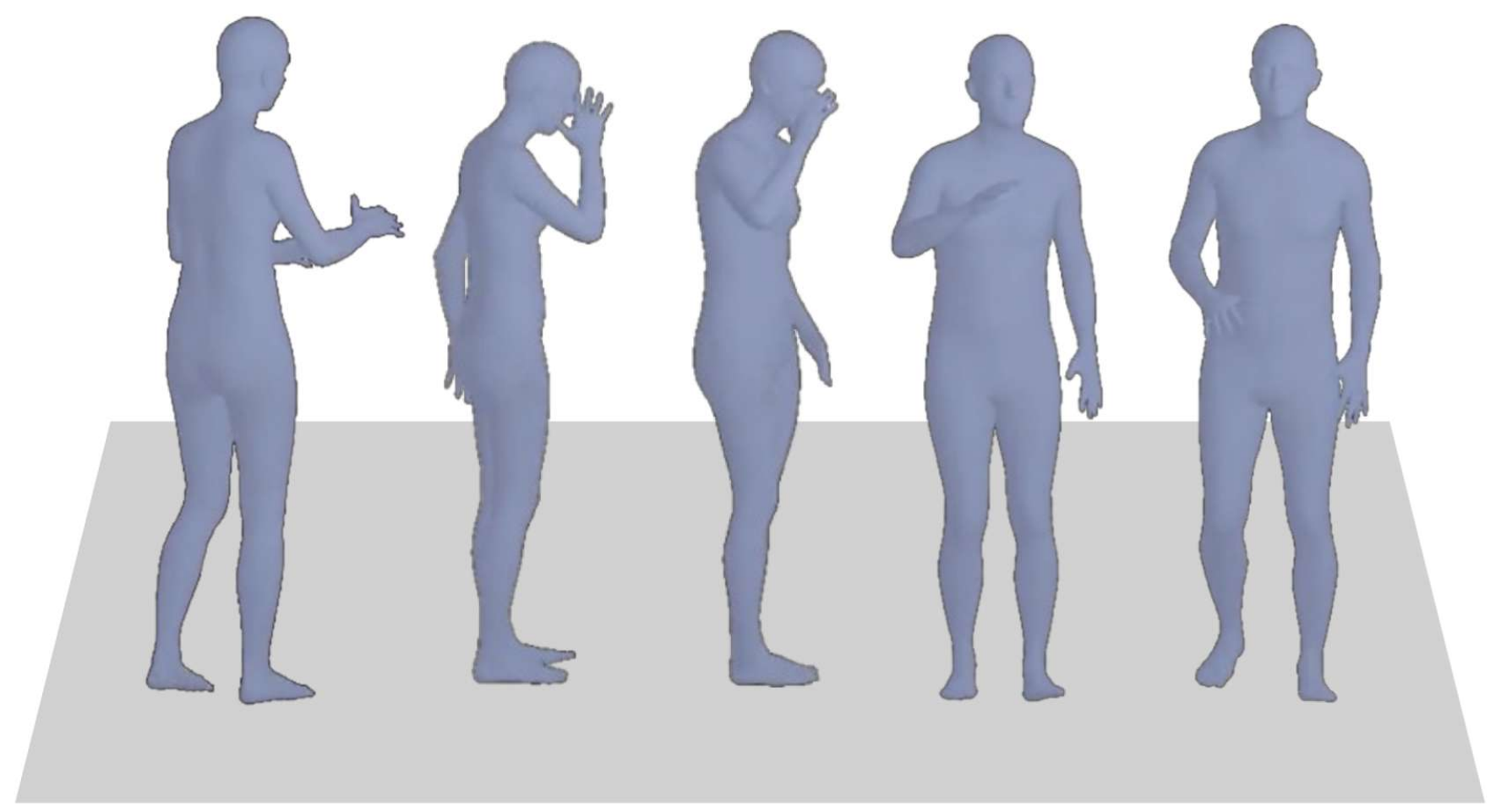}
    \caption{Mesh for Compound Human Actions.}
    \label{fig:small_1}
  \end{subfigure}\hfill
  \begin{subfigure}[b]{0.49\linewidth}
    \centering
    \includegraphics[width=\linewidth]{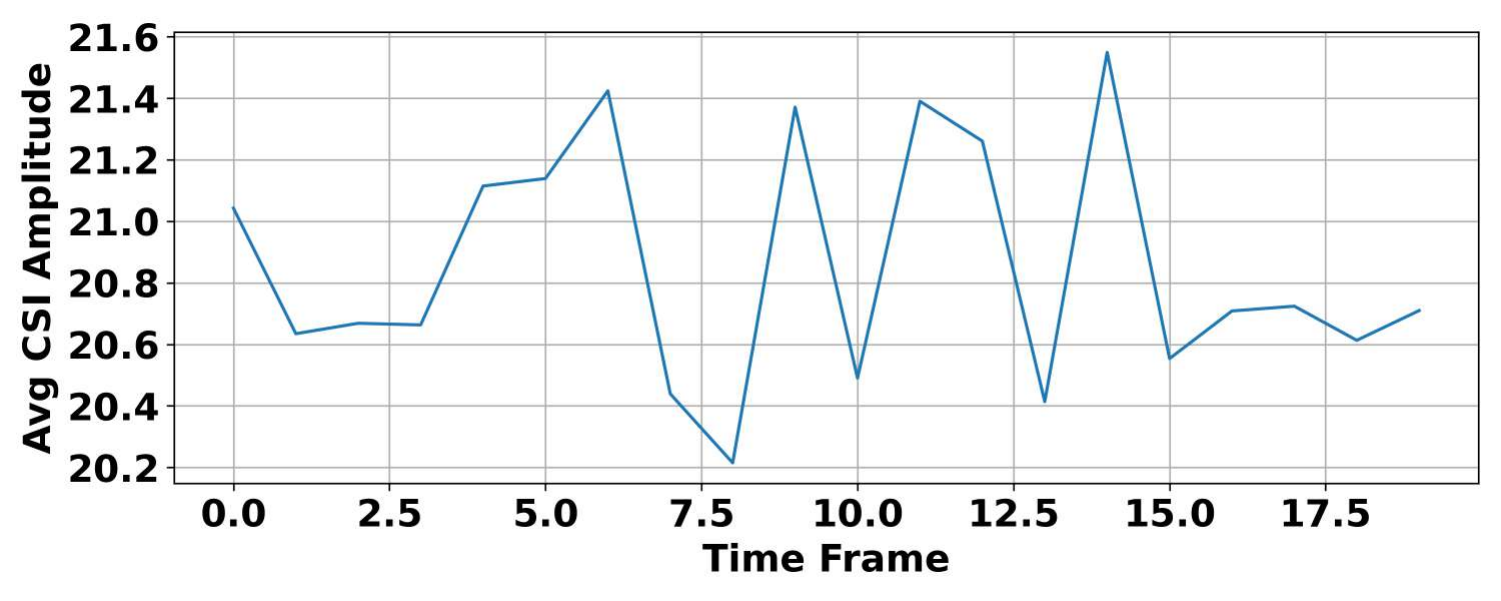}
    \caption{Averaged Amplitude (Unsegmented).}
    \label{fig:small_2}
  \end{subfigure}
  \caption{Visualization of Compound Human Actions and Unsegmented CSI Data.}
  \label{fig:main_figure}
\end{figure}

As illustrated in Fig.~\ref{system_model}, the wireless sensing system from \cite{yan2024person} is equipped with a single transmitter and three receivers, each receiver with three antennas. The transmitter emits wireless signals that traverse the environment and interact with the human bodies performing sequential movements within the sensing area. Fig.~\ref{fig:small_1} illustrates such complex human behavior by demonstrating a sequence of compound activities. The corresponding wireless measurements captured by the receiver array are shown in Fig.~\ref{fig:small_2}, where the CSI amplitude evolves over time in response to these activities. As demonstrated, multiple actions may overlap within the sensing region, leading to a continuous and unsegmented CSI stream without a clear boundary between actions. This observation motivates our formulation to directly operate on the unsegmented CSI sequences, without relying on explicit temporal segmentation.  

To model the underlying CSI signal propagation, we adopt a standard linear channel model at the receiver side and represent the received signal at the $i$th subcarrier as:

\begin{equation}
Y_i = H_iX_i + N_i,
\label{eq:signal-model}
\end{equation}

where \(H_i \in \mathbb{C}^{9\times1}\) denotes the channel matrix, \(X_i\) is the transmitted signal, and with \(N_i\) represents the additive noise.  

\subsection{\textbf{Problem Definition}} 

\textbf{Objective.} Assume our training dataset consists of raw CSI data $Y_i$, derived from the signal model in Eq.~\ref{eq:signal-model}, synchronized video data $X_i$, and natural language annotations $\hat{e}_i$ that describe the corresponding human motions captured in the scene. Since language generation is a supervised task, the annotation set $\hat{e}_i$ is available during the training. Under this setting, the objective of this work is to learn the mapping from the unsegmented CSI measurements to generate an accurate and semantically meaningful natural language description $e'_i$ of human motion directly from continuous wireless observations.      

\begin{figure*}
\centering
\includegraphics[width=0.85\textwidth]{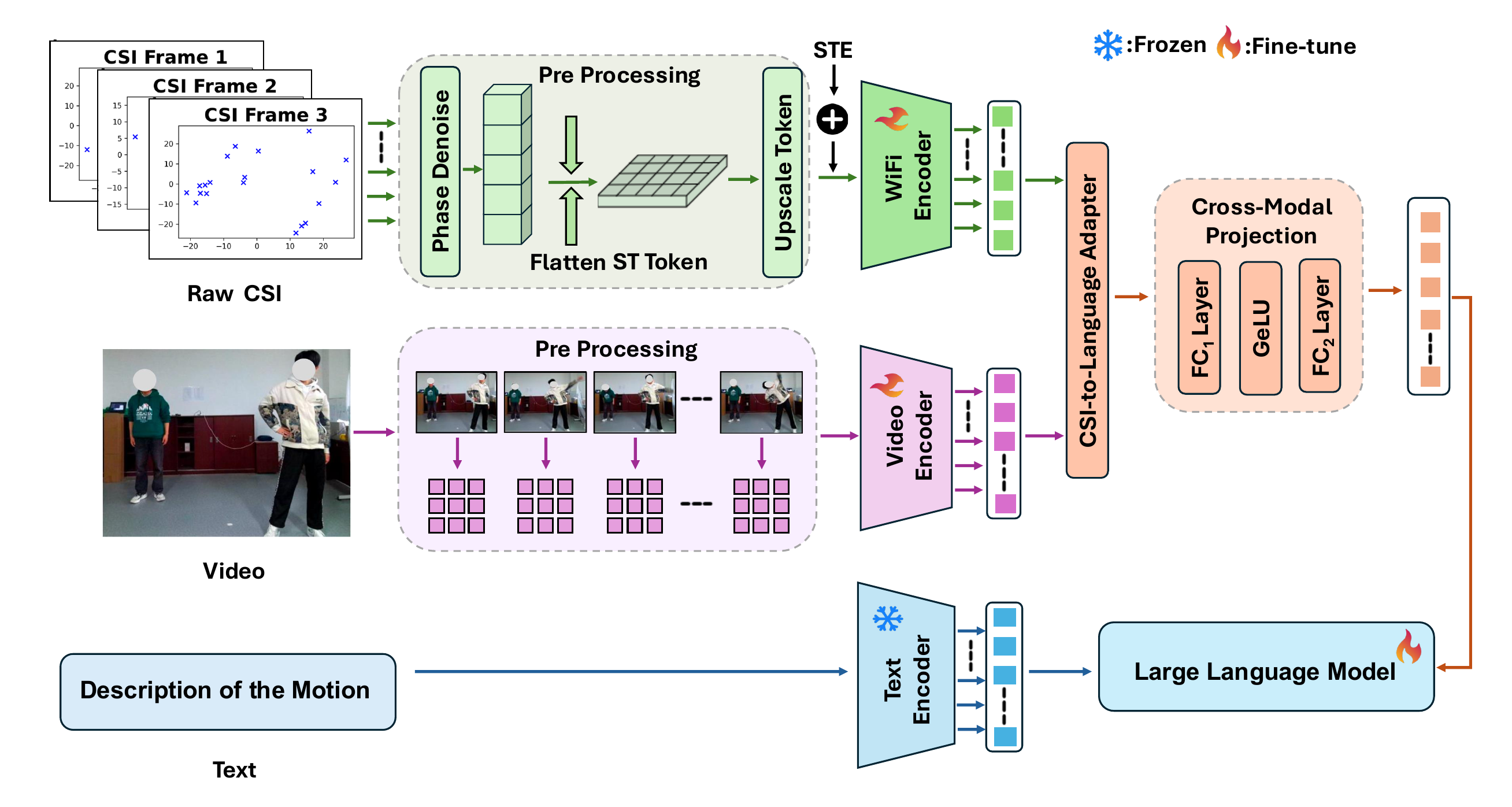}\caption{{WirelessSenseLLM takes raw CSI \(Y_i\) data and text prompts $\hat{e}_i$ as inputs. Synchronized video $X_i$ data is provided only during training as semantic supervision for CSI. The system first processes the CSI data using WiFi Encoder \(E_{csi}\) and video using Video Encoder \(E_v\). In stage 1, the CSI-to-Language Adapter maps the encoded features into language-aligned semantic space $Z_l$. In stage 2, the aligned embeddings are transformed by a cross-modal projection layer into a token representation for LLM and jointly fine-tuned with a large language model for natural language generation. }}
\label{archi}
\end{figure*}

\textbf{Framework.} Fig.~\ref{archi} illustrates the overall framework for transforming raw CSI into natural language descriptions. Our framework consists of a WiFi encoder $E_{csi}$, a video encoder $E_v$, a text encoder $E_{txt}$, a CSi-to-Language Adapter, a cross-modal projection module $f_p$, and a large language model $f_l$. The WiFi encoder $E_{csi}$ takes the raw CSI $Y_i$ as input and extracts latent wireless embeddings $f_{csi}^i$ from the continuous and unsegmented CSI streams. Meanwhile, the video encoder $E_v$ takes the synchronized video frames $X_i$ as input and generates video embeddings $f_v^i$, which serve as semantic supervision during the training. These CSI embeddings $f_{csi}^i$ are subsequently aligned using the CSI-to-Language adapter, resulting in $f_{align}^i$ and then transformed using the cross-modal projection $f_p$ into an LLM-compatible token. Additionally, $E_{txt}$ is used to encode the textual prompt into prompt embeddings $f_{txt}^i$, guiding the language generation process. After the projection, embeddings are concatenated with the textual prompts and provided as input to the language model $f_l$. The output of the language model is a token sequence: 

\begin{equation}
 Z, Z^* = \{s_t, s^*_t\}_{t=1}^{T}     
\end{equation}

where $Z = \{s_t \}_{t=1}^T$ denotes the token sequence generated by the language model from language description $e'_i$ and $Z^* = \{s_t^* \}_{t=1}^T$ denotes the ground truth token sequence from annotations $\hat{e}_i$. The model is trained to maximize the likelihood of generating the ground truth annotations $\hat{e}_i$ in an auto-regressive manner, which is achieved by minimizing the cross-entropy loss as: 

\begin{equation}
L = - \sum_{k=1}^{L} \log P\left( s^{*}_k \mid s^{*}_{<k}, f_{p}'; \theta \right)
\end{equation}

where $s^*_{<k}$ denotes the ground truth tokens that are previously generated, and $\theta$ represents the trainable parameters of the model.

\section{Methodology}\label{AA}
\label{sec:four}

\textbf{Training Pipeline}

The training pipeline of the WirelessSenseLLM is divided into two stages. In the first stage, the model performs a semantic pre-alignment using the CSI-to-Language Adapter, where CSI embeddings are aligned with language semantics while utilizing the synchronized video as additional supervision. In the second stage, the aligned CSI embeddings are adapted for LLM to enable the natural language description.


\textbf{CSI-to-Language Adapter (Stage 1).} Adopting the CSI representation and tokenization method from \cite{yan2024person}, each sample is first converted into a spatio-temporal token sequence by concatenating amplitude and phase information along the subcarrier dimension, resulting in an input tensor of $Y_i \in \mathbb{R}^{180 \times 60}$. These CSI tokens are projected to match the LLM model embedding dimension using a fully connected layer, resulting in $(Y_i \in \mathbb{R}^{180 \times 256})$. Learnable spatio-temporal embeddings (STE) are then added to preserve the temporal structure across CSI tokens, resulting in the final encoder input of $Y_{st}^i \in \mathbb{R}^{180 \times 256}$. This sequence is then processed by the WiFi encoder following the backbone in \cite{yan2024person} to generate refined features $f_{enc}^i$ that capture motion-induced temporal patterns. However, these features remain domain-specific and are not directly compatible with the language domain. To address this, the CSI-to-Language Adapter performs a pre-alignment step using a two-layer MLP with GeLU activation to map CSI embeddings into language-aligned semantic space $Z_l$. The video embedding already resides in the language space and acts as an additional semantic supervision during training. 

\begin{figure}[]
  \begin{center}
\includegraphics[width=0.5\textwidth]{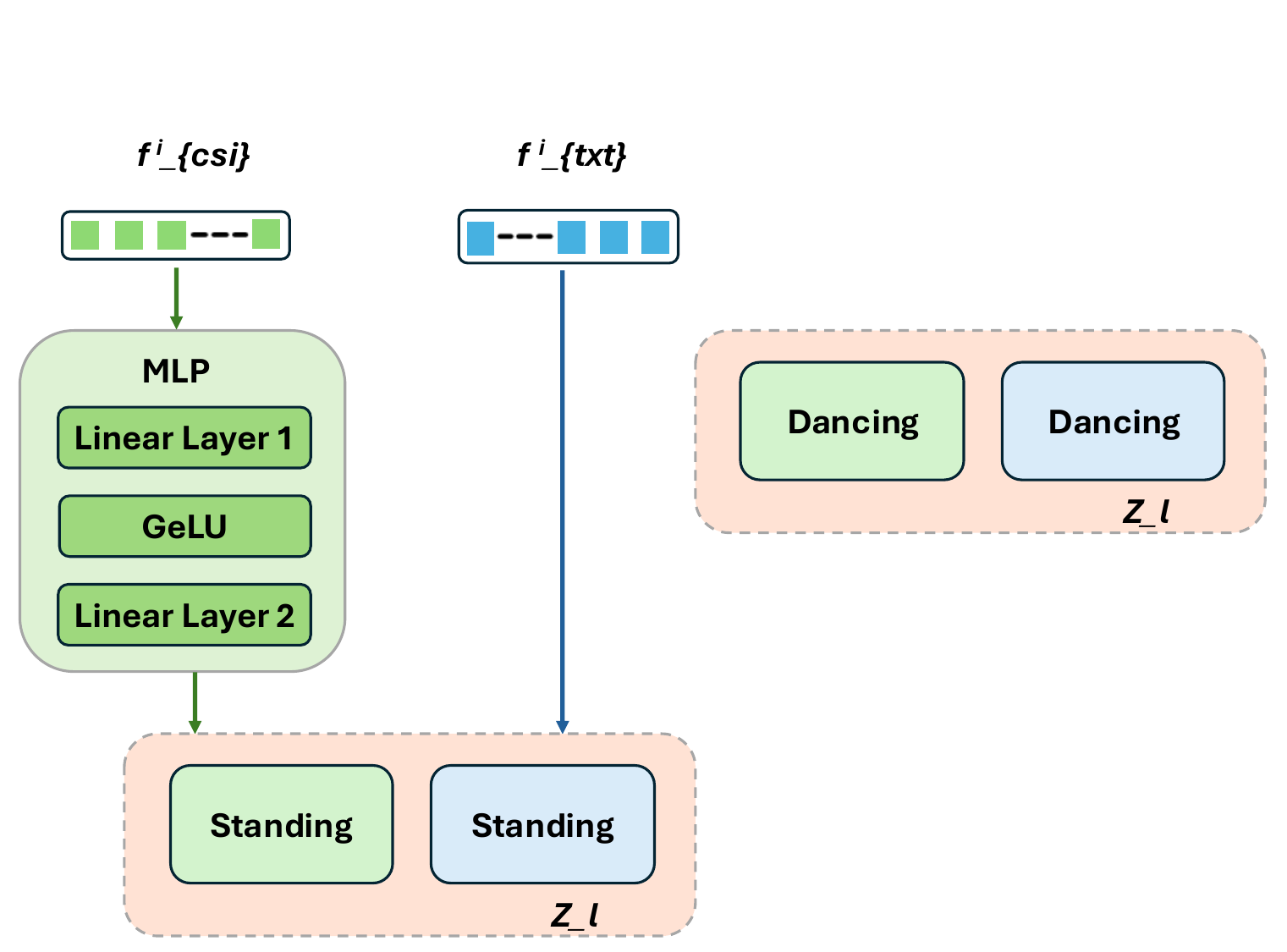}
    \caption{CSI embeddings are contrastively aligned with frozen text embedding in a shared language space $Z_l$ to bridge the CSI to Language modality gap.}
    \label{pre_align}
    \end{center}
\end{figure}

The text encoder $E_{txt}$ processes the corresponding language annotations derived from synchronized video into the shared semantic space $Z_l$ and remains frozen. Contrastive learning is then applied to align the CSI embeddings with their corresponding textual embeddings by maximizing the similarity between CSI-text pairs and minimizing the similarity for mismatched pairs (Fig.~\ref{pre_align}), with the loss defined as: 
\begin{equation}
    L = L_{c2t} + L_{t2c}    
\end{equation}

where $L_{c2t}$ denotes the CSI-to-text contrastive loss and $L_{t2c}$ denotes the text-to-CSI contrastive loss. During this stage, only the CSI-to-Language adapter is trained, and all other components remain frozen. The video encoder is also kept frozen throughout this stage, as its outputs are already aligned with language semantics. This adapter design effectively resolves the representation gap between the continuous CSI signals and discrete language tokens. Moreover, it operates directly on the unsegmented CSI sequences and preserves the temporal structure for modeling overlapping and sequential human actions.  

\begin{figure}[h!]
    \centering
    \begin{subfigure}[b]{0.49\columnwidth}
        \centering
        \includegraphics[width=4.5cm]{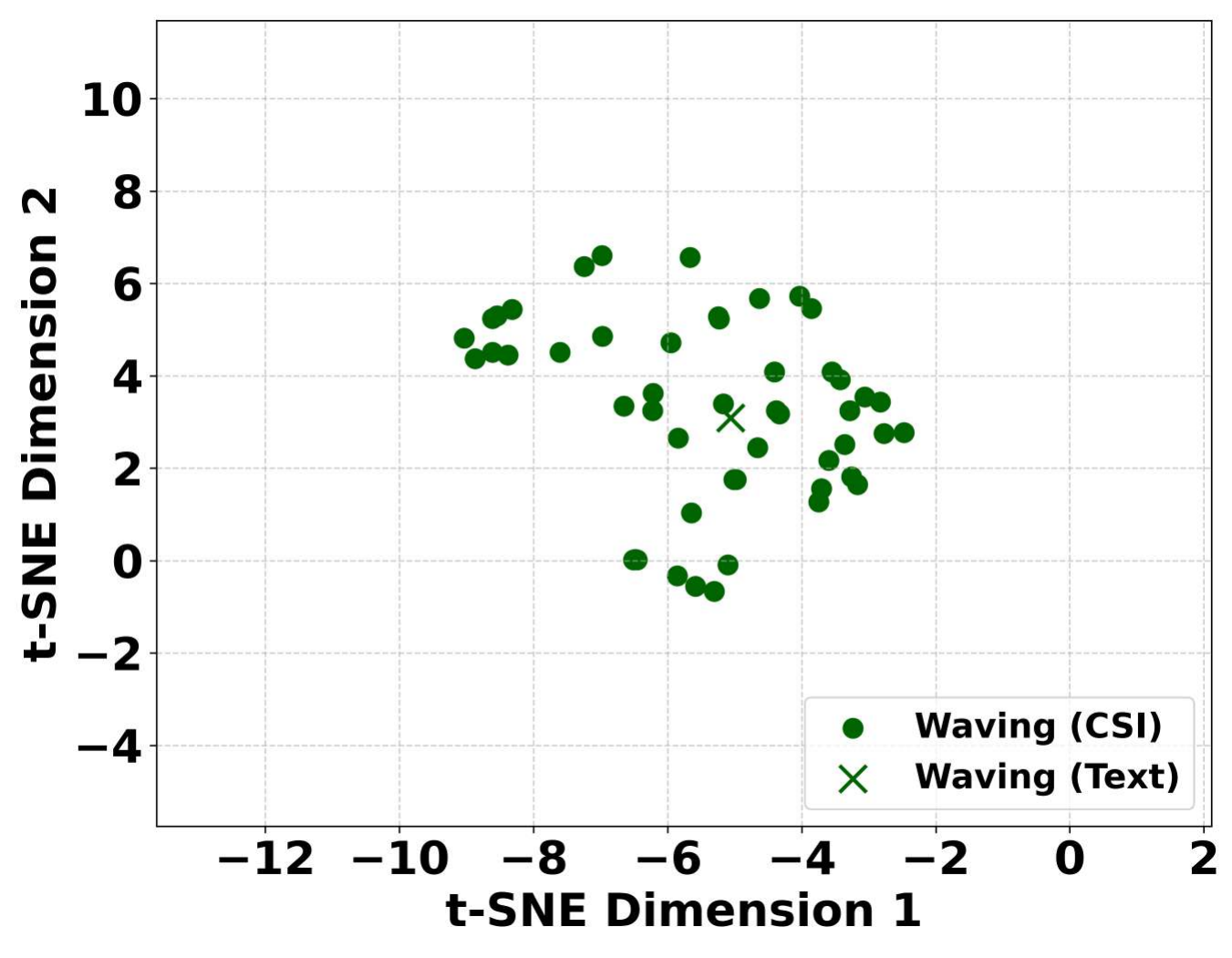}
        \subcaption{}
        \label{fig:single_p_tsne}
    \end{subfigure}
    \hfill
    \begin{subfigure}[b]{0.49\columnwidth} 
        \centering
        
        \includegraphics[width=4.5cm]{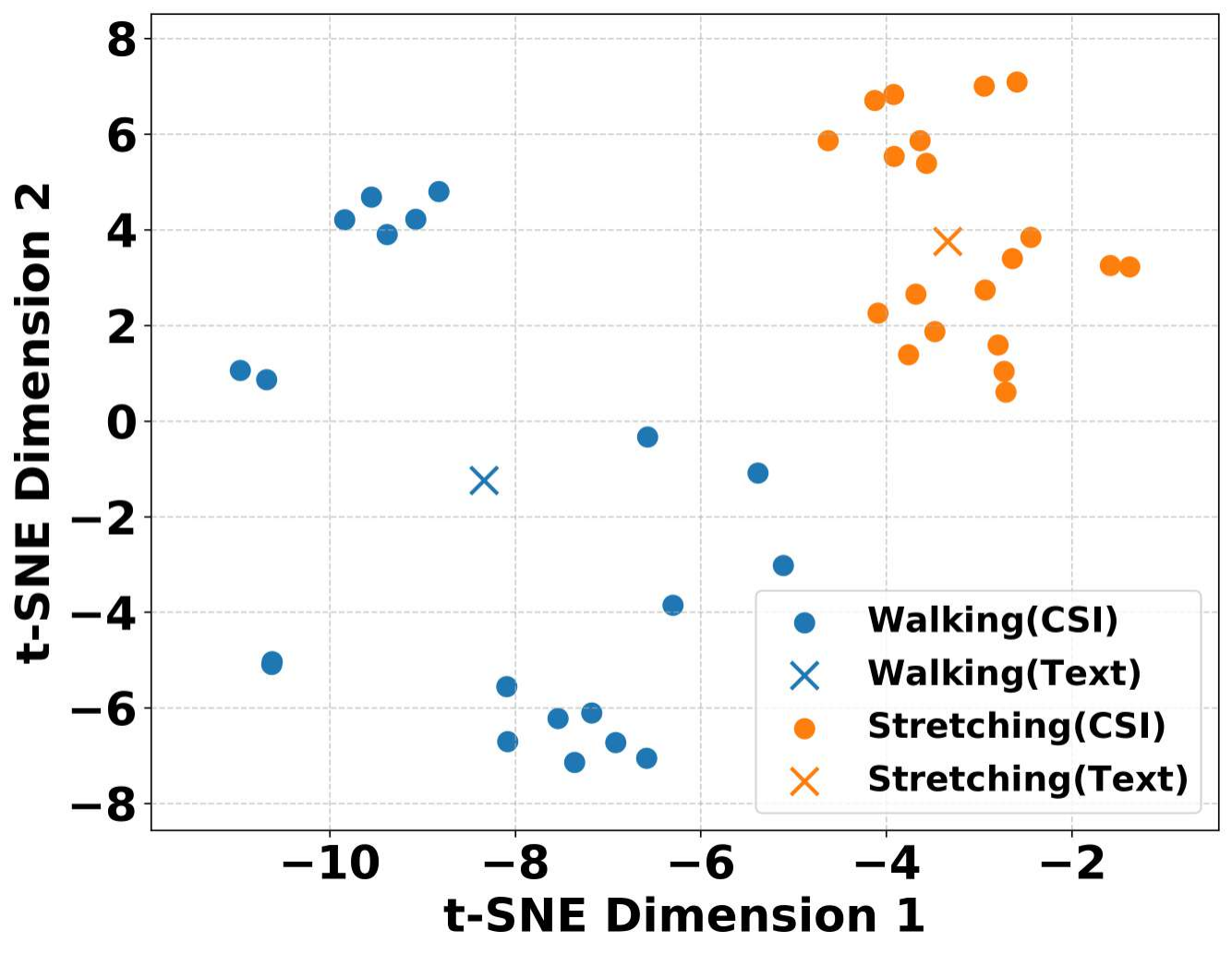}
        \subcaption{}
        \label{fig:multi_p_tsne}

    \end{subfigure}
    \caption{Projection Layer performance of WirelessSenseLLM across Single Person and Two Person Scenario.}
    \label{fig:t-sne}
\end{figure}

\begin{figure}[]
  \begin{center}
\includegraphics[width=0.50\textwidth]{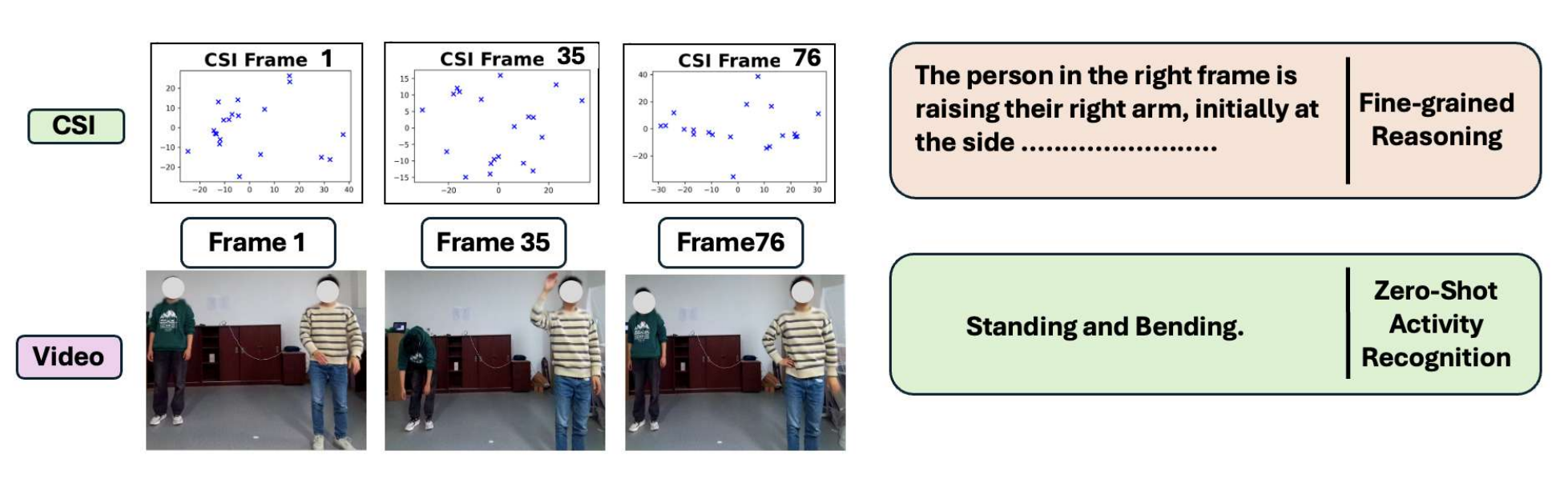}
    \caption{WirelessSenseLLM example for Two Persons.}
    \label{example_two}
    \end{center}
\end{figure}

\textbf{Cross-Modal Projection and LLM Fine-Tuning (Stage 2).} In the second stage, the aligned CSI embeddings $f_{align}^i$ from the CSI-to-Language Adapter are transformed into token-level representations that are compatible with the large language model. For this purpose, the aligned CSI embeddings and their corresponding video embeddings $f_{v}^i$ are concatenated to form a unified representation $f_{uni}^i$ to insert into the cross-modal feature projection layer $f_p$ to generate language-compatible token-level inputs. We can define the $f'_p$ as:

\begin{equation}
 f_p' = [f_{align}^i ; f_{v}^i]
\end{equation}

These projected tokens are then subsequently provided to the LLM model $f_l$ for instruction tuning. During this stage, the LLM and the projection layer are jointly fine-tuned to generate instruction-driven, fine-grained natural language responses for human motion captured from the unsegmented and continuous CSI streams. To enable parameter-efficient adaptation while preserving the pre-trained language knowledge, Low-Rank Adaptation (LoRA) \cite{hu2022lora} is applied to the LLM.

The effectiveness of the learned representations after the projection layer is summarized in Fig.~\ref{fig:t-sne}, with (a) showing that text embeddings and wireless embeddings for a single person scenario clustered together, and (b) demonstrating a clear separation between different actions while still maintaining the close alignment between corresponding wireless embeddings and text embeddings in multi-person scenarios. As further illustrated in Fig.~\ref{example_two}, after the instruction tuning, the model can generate descriptive and interpretable natural language that explains different human motions directly from unsegmented CSI data.

\section{Experiments}\label{AA}  

\subsection{\textbf{Dataset Preparation:}}
To the best of our knowledge, no publicly available CSI dataset is specifically annotated for LLM under multi-person scenarios. Existing wireless sensing datasets such as CSI-bench \cite{zhu2025csi}, WiMANS \cite{huang2024wimans}, UT-HAR \cite{yousefi2017survey}, Widar \cite{yang2020widar}, and NTU-Fi HAR \cite{yang2023sensefi}, are largely designed for motion detection or activity recognition. As a result, they lack the natural language annotations required for LLM-based reasoning. To address this limitation, inspired by \cite{chen2024motionllm, lin2023video}, we introduce a wireless sensing dataset with natural language annotations. The dataset includes three types of paired data: Wireless-Text, Video-Text, and Text-only annotations.  

\paragraph{Wireless Dataset} We adopt the CSI dataset from \cite{yan2024person}, in which multiple volunteers performed simultaneous activities such as raising hands, sitting down, and lifting legs. During data collection, the CSI receivers and the camera were manually synchronized. The video is recorded at 15 fps, and the CSI transmission rate was 300 packets per second. Based on this alignment, each video frame corresponds to 20 CSI packets. To construct the text supervision, initial captions are generated using LLaVA Next \cite{li2024llava} with carefully designed prompts, and subsequently refined using GPT-4 to improve the spatio-temporal description quality and reasoning consistency. All annotations are then manually reviewed by a human to ensure correctness with the underlying CSI data.  

\paragraph{Video Pair}
WirelessSenseLLM leverages video data as ground truth semantic supervision during training to guide the representation learning of CSI. The video dataset is sourced from both MoVid \cite{chen2024motionllm} and Person-in-WiFi 3D \cite{yan2024person}, which provide the rich natural-language annotations describing diverse human motions. In total, there are 24k video annotation pairs, which serve as additional semantic reference during training. 

\paragraph{Text} Text-Only annotations are used during the cross-modal feature projection stage to align the projected tokens with language representations. The dataset is also sourced from the MoVid \cite{chen2024motionllm} with 40k text entries. Summary of the dataset can be found in Table~\ref{tab:csi_dataset_stat}. 

\setlength{\tabcolsep}{10pt}
\renewcommand{\arraystretch}{1.2}

\begin{table}[h!]
\centering
\begin{tabularx}{0.47\textwidth}{|>{\centering\arraybackslash}X | >{\centering\arraybackslash} X | >{\centering\arraybackslash} X | >{\centering\arraybackslash} X |}
\hline
Dataset & Wireless-Text & Video-Text & Text \\
\hline
Train & 89946 & 22588 & 2510 \\
\hline
Test & 7824 & 36619 & 4069 \\
\hline
In Total & 97000 & 25098 & 40688
\\
\hline
\end{tabularx}

\caption{\textbf{WirelessSenseLLM} Dataset Overview}  
\label{tab:csi_dataset_stat}
\end{table}

\subsection{\textbf{Comparative Analysis}}
In this section, we present the comparative analysis and visualization of our model's performance. We evaluate on two fundamental tasks: Zero-Shot Human Motion Understanding and In-Depth Reasoning from CSI, particularly under settings where explicit temporal segmentation and action-specific labels are unavailable. Each of these tasks is analyzed in single-person and multi-person scenarios. The evaluation reports multiple automated scoring metrics, e.g., METEOR, ROUGE-1, ROUGE-L, BLEU, and BERTScore, with LLM-driven judgments. It will deliver both quantitative and qualitative insights into model performance.

\newcolumntype{Y}{>{\hsize=1.0\hsize\arraybackslash}X}
\setlength{\tabcolsep}{10pt}
\renewcommand{\arraystretch}{1.2}

\begin{table}[h!]
\centering
\begin{tabularx}{0.47\textwidth}{|>{\centering\arraybackslash}Y|>{\centering\arraybackslash}Y|>{\centering\arraybackslash}Y|}
\hline
Method & Accuracy & F1 Score \\
\hline
SVM & 0.29 & 0.25 \\
\hline
Vision SVM & 0.14 & 0.13 \\
\hline
Vision CNN & 0.37 & 0.30 \\
\hline
Vision RNN & 0.40 & 0.33 \\
\hline
WirelessSenseLLM & 0.67 & 0.65 \\
\hline
WirelessSenseLLM + ICL & 0.89 & 0.87 \\
\hline
WirelessSenseLLM + COT & 0.92 & 0.91 \\
\hline
\end{tabularx}
\caption{Comparison between WirelessSenseLLM and Traditional Models on Zero-Shot Human Actions}
\label{tab:zero_shot}
\end{table}

First, we evaluate our model's performance on zero-shot human action understanding using standard classification metrics, including accuracy and F1 score. In this work, zero-shot refers to evaluation on actions that are excluded from the training. 

As shown in Table~\ref{tab:zero_shot}, traditional baselines including SVM, Vision SVM, Vision CNN, and Vision RNN, exhibit limited performance, with accuracy ranging from 14\% to 40\% and F1-scores between 13\% and 33\%. It is important to note that these baselines are evaluated on well-segmented wireless signals, and yet still struggle to generalize to unseen activities because they heavily rely on segmented inputs and pre-defined labeled activity classes. This makes them ineffective for zero-shot evaluation on unsegmented CSI data. In contrast, WirelessSenseLLM clearly outperforms the compared baselines, achieving $67\%$ accuracy and a $65\%$ F1 score with a basic prompting strategy. Such improvement demonstrates the model's ability to infer human actions directly from unsegmented CSI by utilizing the language-level reasoning. When enhanced with In-Context Learning (ICL), where a few example pairs are given in the prompt, performance further improves to $89\%$ accuracy and $87\%$ F1 score. This indicates that contextual examples help LLM to interpret motion semantics better. Integrating chain-of-thoughts (CoT) \cite{wei2022chain} reasoning, which allows the LLM to reason through intermediate steps before the final prediction, yields a further gain to $92\%$ accuracy and $91\%$ F1 score. Overall, these results confirm that our approach effectively addresses the key limitation of the prior work by operating without explicit segmentation and without action-specific training labels. 

\begin{table*}[h!]
\centering

\renewcommand{\arraystretch}{1.2}
\begin{tabular}{|l|cc|cc|cc|cc|cc|cc|}
\hline
\textbf{Model} & \multicolumn{2}{|c|}{\text{Fac.}} & \multicolumn{2}{|c|}{\text{Temp. Flow}} & \multicolumn{2}{|c|}{\text{Spa. Rel.}} & \multicolumn{2}{|c|}{\text{Body P.}} & \multicolumn{2}{|c|}{\text{Int.}} & \multicolumn{2}{|c|}{\text{All}}\\
\cline{2-13}

& Acc. & Score & Acc. & Score & Acc. & Score & Acc. & Score & Acc. & Score & Acc. & Score \\

\hline
GT (1 P) & 100.0 & 5.00 & 100.0 & 5.00 & 100.0 & 5.00 & 100.0 & 5.00 & 100.0 & 5.00 & 100.0 & 5.00 \\

\text{\cite{lin2023video}} & \text{16.67} & \text{1.12} & \text{31.25} & \text{1.67} & \text{0.0} & \text{0.29} & \text{4.17} & \text{0.62} & \textbf{91.67} & \textbf{2.08} & \text{26.85} & \text{1.62} \\

\textbf{Our} & \textbf{75.00} & \textbf{2.96} & \textbf{39.58} & \textbf{2.31} & \textbf{29.17} & \textbf{1.75} & \textbf{8.33} & \textbf{1.42} & \text{29.17} & \text{1.21} & \textbf{33.33} & \textbf{1.87} \\

\hline

\end{tabular}
\label{tab1}
\caption{Evaluation with baseline for Single person. WirelessSenseLLM surpasses the baseline in overall average metrics.}
\label{tab:llm_category}
\end{table*}

\begin{table*}[h!]
\centering

\renewcommand{\arraystretch}{1.2}

\begin{tabular}{|p{3.5cm}|p{3.5cm}|c|c|c|c|c|}
\toprule
\textbf{Model} & \textbf{Category} & \text{ROUGE-1} & \text{ROUGE-L} &\text{BLEU} & \text{METEOR} & \text{BERTScore}\\
\hline

\multirow{7}{*}{\textbf{Video-LLaVA}}
& Overall & 0.2996 & 0.1999 & 0.0284 & 0.1671 & 0.8638 \\
& Factuality & 0.3290 & 0.2138 & 0.0353 & 0.1684 & 0.8613 \\
& Temporal Flow & 0.2929 & 0.1883 & 0.0140 & 0.1726 & 0.8669 \\
& Spatial Rel. & 0.3729 & 0.2998 & 0.1011 & 0.2422 & 0.8885 \\
& Body Part & 0.2218 & 0.1468 & 0.0019 & 0.0857 & 0.8328 \\
& Interaction & 0.3700 & 0.2175 & 0.0590 & 0.2166 & 0.8734 \\
& Action Identification & 0.2320 & 0.1668 & 0.0035 & 0.1008 & 0.8506 \\
\midrule

\multirow{7}{*}{\textbf{WirelessSenseLLM}}
& Overall & 0.3352 & 0.2376 & 0.0379 & 0.1937 & 0.8800 \\
& Factuality & 0.3457 & 0.2782 & 0.0505 & 0.1937 & 0.8815 \\
& Temporal Flow & 0.3671 & 0.2556 & 0.0556 & 0.2211 & 0.8934 \\
& Spatial Rel. & 0.3622 & 0.2752 & 0.0742 & 0.2386 & 0.8905 \\
& Body Part & 0.3013 & 0.1977 & 0.0100 & 0.1301 & 0.8527 \\
& Interaction & 0.2877 & 0.1935 & 0.0258 & 0.1678 & 0.8654 \\
& Action Identification & 0.2451 & 0.1725 & 0.0059 & 0.1277 & 0.8568 \\
\hline

\end{tabular}
\label{tab1}
\caption{Average scoring for One Person Scenario.} WirelessSenseLLM outperforms the baseline.
\label{tab:traditional_nlp_evaluation}
\end{table*}

\begin{table*}[h!]
\centering
\renewcommand{\arraystretch}{1.2}
\begin{tabular}{|l|cc|cc|cc|cc|cc|cc|}
\hline
\textbf{Model} & \multicolumn{2}{|c|}{\text{Fac.}} & \multicolumn{2}{|c|}{\text{Flow}} & \multicolumn{2}{|c|}{\text{Rel.}} & \multicolumn{2}{|c|}{\text{Body}} & \multicolumn{2}{|c|}{\text{Int.}} & \multicolumn{2}{|c|}{\text{All}}\\
\cline{2-13}

& Acc. & Score & Acc. & Score & Acc. & Score & Acc. & Score & Acc. & Score & Acc. & Score \\

\hline
GT (2 P) & 100.0 & 5.00 & 100.0 & 5.00 & 100.0 & 5.00 & 100.0 & 5.00 & 100.0 & 5.00 & 100.0 & 5.00 \\  

\text{\cite{lin2023video}} & \textbf{87.83} & \textbf{4.00} & \text{28.26} & \text{1.37} & \text{0.00} & \text{0.65} & \text{0.00} & \text{1.50} & \text{6.52} & \textbf{1.04} & \text{26.52} & \text{1.33} \\

\textbf{Our} & \text{45.65} & \text{1.35} & \textbf{58.70} & \textbf{1.65} 
& \textbf{28.26} & \textbf{1.43} & \textbf{58.70} & \textbf{1.65} 
& \textbf{23.91} & \text{0.93} & \textbf{44.78} & \textbf{1.47} \\

\midrule\midrule

GT (3 P) & 100.0 & 5.00 & 100.0 & 5.00 & 100.0 & 5.00 & 100.0 & 5.00 & 100.0 & 5.00 & 100.0 & 5.00 \\  

\text{\cite{lin2023video}} & \textbf{71.74} & \textbf{3.50} & \text{2.17} & \text{1.28} & \text{0.00} & \text{0.22} & \text{0.00} & \text{1.13} & \text{26.09} & \textbf{2.07} & \text{20.00} & \text{1.10} \\

\textbf{Our} & \text{19.57} & \text{0.89} & \textbf{45.65} & \textbf{1.17} 
& \textbf{34.78} & \textbf{1.33} & \textbf{36.96} & \textbf{1.28} 
& \textbf{41.30} & \text{1.28} & \textbf{35.65} & \textbf{1.19} \\

\midrule\midrule

GT (4 P) & 100.0 & 5.00 & 100.0 & 5.00 & 100.0 & 5.00 & 100.0 & 5.00 & 100.0 & 5.00 & 100.0 & 5.00 \\  

\text{\cite{lin2023video}} & \textbf{53.85} & \textbf{2.85} & \text{0.00} & \text{0.92} & \text{7.69} & \text{0.77} & \text{0.00} & \text{0.13} & \textbf{30.77} & \textbf{1.92} & \text{18.46} & \text{1.00} \\

\textbf{Our} & \text{30.77} & \text{1.15} & \textbf{23.08} & \textbf{0.92} 
& \textbf{30.77} & \textbf{1.15} & \textbf{2.00} & \textbf{0.69} 
& \text{23.08} & \text{1.15} & \textbf{21.54} & \textbf{1.02} \\

\hline

\end{tabular}
\label{tab1}
\caption{Evaluation with baseline for two, three, and four persons. WirelessSenseLLM surpasses the baseline in all scenarios.}
\label{tab:multi_category}
\end{table*}

\subsection{\textbf{Language-driven Motion Understanding and Reasoning}}

In this section, we evaluate the model's performance on generating accurate and semantically meaningful natural language descriptions from the CSI in both single-person and multi-person scenarios.  

\paragraph{\textbf{Single-Person Results}} For a single person, we utilized two evaluation protocols: (1) LLM-based judgment using GPT-4o as an external evaluator, and (2) automated text similarity metrics.  

\textbf{GPT-4o as judge:} Our GPT-4o-as-judge evaluation protocol is inspired by \cite{lin2023video} and is designed to assess the reasoning quality of generated descriptions for sequential complex movements against the ground truth annotations. The evaluation focuses on five aspects: factual correctness, temporal flow, spatial relationships, body part descriptions, and interaction understanding. GPT-4o assigns scores on a scale from 0 to 5, where higher scores indicate stronger reasoning quality and closer alignment with the ground truth. 
 
Table~\ref{tab:llm_category} summarizes the evaluation results for Video-LLaVA \cite{lin2023video} and our proposed WirelessSenseLLM. Video-LLaVA exhibits uneven performance across all five categories, achieving relatively strong results primarily in interaction-related descriptions while showing limited capability in factual, temporal, and spatial reasoning. In contrast, WirelessSenseLLM demonstrates more consistent improvement across most categories, despite relying solely on wireless signals. Specifically, our model achieves $75\%$ accuracy in factual descriptions, and $39.58\%$ in temporal flow, outperforming the baseline by a substantial margin. Similar to Video-LLaVA, performance remains lower for spatial relationships and fine-grained body part recognition. This highlights the inherent challenge in recovering precise spatial anatomy from CSI alone. Collectively, our model's average accuracy improved by approximately $30\%$, and the reasoning score by around 15\% relative to the baseline.

\textbf{Automated scoring methods:}

We further utilized automated scoring metrics to assess WirelessSenseLLM's reasoning quality, including: ROUGE-1 and ROUGE-L (Recall-Oriented Understudy for Gisting Evaluation) \cite{lin2004rouge}, BLEU (Bilingual Evaluation Understudy) \cite{papineni2002bleu}, METEOR (Metric for Evaluation of Translation with Explicit Ordering) \cite{banerjee2005meteor}, and BERTScore \cite{zhang2019bertscore}. These methods collectively measure content coverage, lexical precision, semantic similarity, and contextual alignment between the generated descriptions and ground truth annotations. 

As summarized in Table~\ref{tab:traditional_nlp_evaluation}, category-level results show that WirelessSenseLLM achieves the most significant gains in factual description and temporal reasoning. Notably, ROUGE-1 for factuality improves by approximately $5\%$, and over $30\%$ in ROUGE-L, highlighting accurate identification of actions. Temporal flow exhibits even larger improvements, with ROGUE-1 increasing by more than $25\%$ and BLUE exhibiting a substantial relative increase, demonstrating improved action ordering and motion continuity.

Alongside category-specific improvements, our model also demonstrates consistent gains across all automated evaluation metrics. Specifically, it improves ROUGE-1 by $12\%$ and ROUGE-L by nearly $20\%$ relative to Video-LLaVA. This indicates a better coverage of important motion details. These gains are also reflected in higher METEOR and BERTScore, which capture a strong semantic alignment and contextual coherence. Finally, achievement in BLEU further suggests our model's improved fluency and phrase-level accuracy. 

Overall, these results demonstrate that WirelessSenseLLM successfully addresses the key challenges in wireless motion understanding by eliminating the need for explicit temporal segmentation and action-specific labels. Moreover, with the help of language-level reasoning, our model achieves strong zero-shot understanding and consistent motion description directly from CSI data.

\paragraph{\textbf{Multi-Person Results}} To evaluate our model in a multi-person scenario for reasoning capability, we are also utilizing GPT-4o as an external judge and automated scoring metrics for qualitative comparison, following the same evaluation metrics used for the single-person scenario.  

\textbf{GPT-4o as judge:} Table~\ref{tab:multi_category} summarizes the evaluation results for two, three, and four-person scenarios. In the two-person setting, the baseline attains high factual accuracy of $87.83\%$ but degrades substantially in temporal flow, spatial, and body-part reasoning. This results in an overall accuracy of $26.52\%$. On the other hand, our model achieves more balanced performance across all categories with an overall accuracy of $44.78\%$. This corresponds to a $18.26\%$ improvement over the baseline. Our model also achieves a higher average reasoning score ($1.47$ vs $1.33$). As the number of subjects increases to three, the performance declines for both models due to increased motion overlap and interaction complexity. Even so, our model consistently outperforms the baseline by $15.65\%$ improvement in accuracy, and also the average reasoning score improved from 1.10 to 1.19. Finally, for the four-person scenario, although accuracy drops substantially for both, WirelessSenseLLM continues to provide a modest accuracy improvement of $3.08\%$ while slightly increasing the average reasoning score from $1.00$ to $1.02$.

\textbf{Automated scoring methods:}

\begin{figure}[t]
    \centering
    \begin{subfigure}[b]{0.44\columnwidth}
        \centering
        \includegraphics[width=4.4cm]{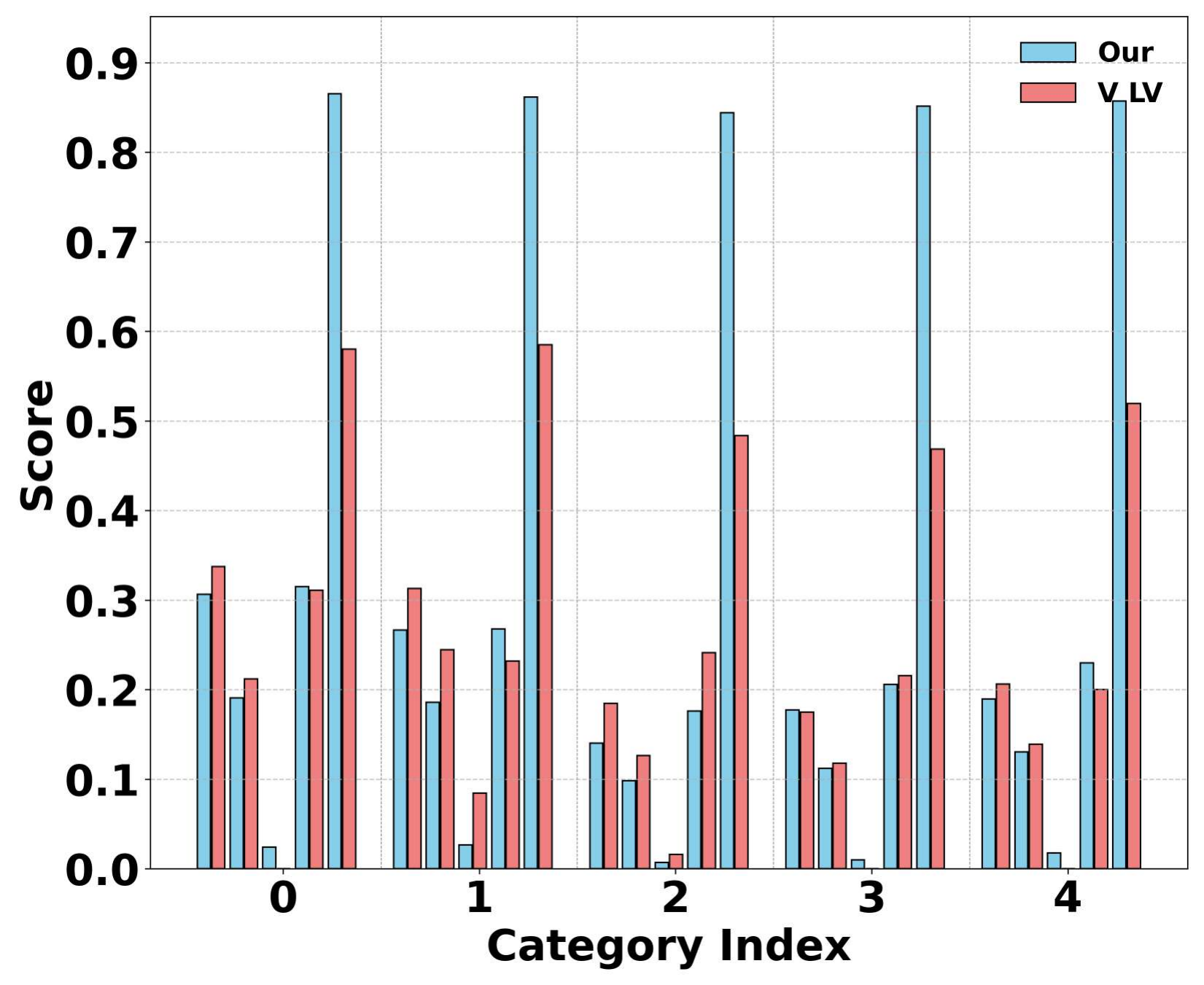}
        \subcaption{Two-person scenario}
        \label{fig:2_person}
    \end{subfigure}
    \hspace{0.5em}
    \begin{subfigure}[b]{0.44\columnwidth}
        \centering
        \includegraphics[width=4.4cm]{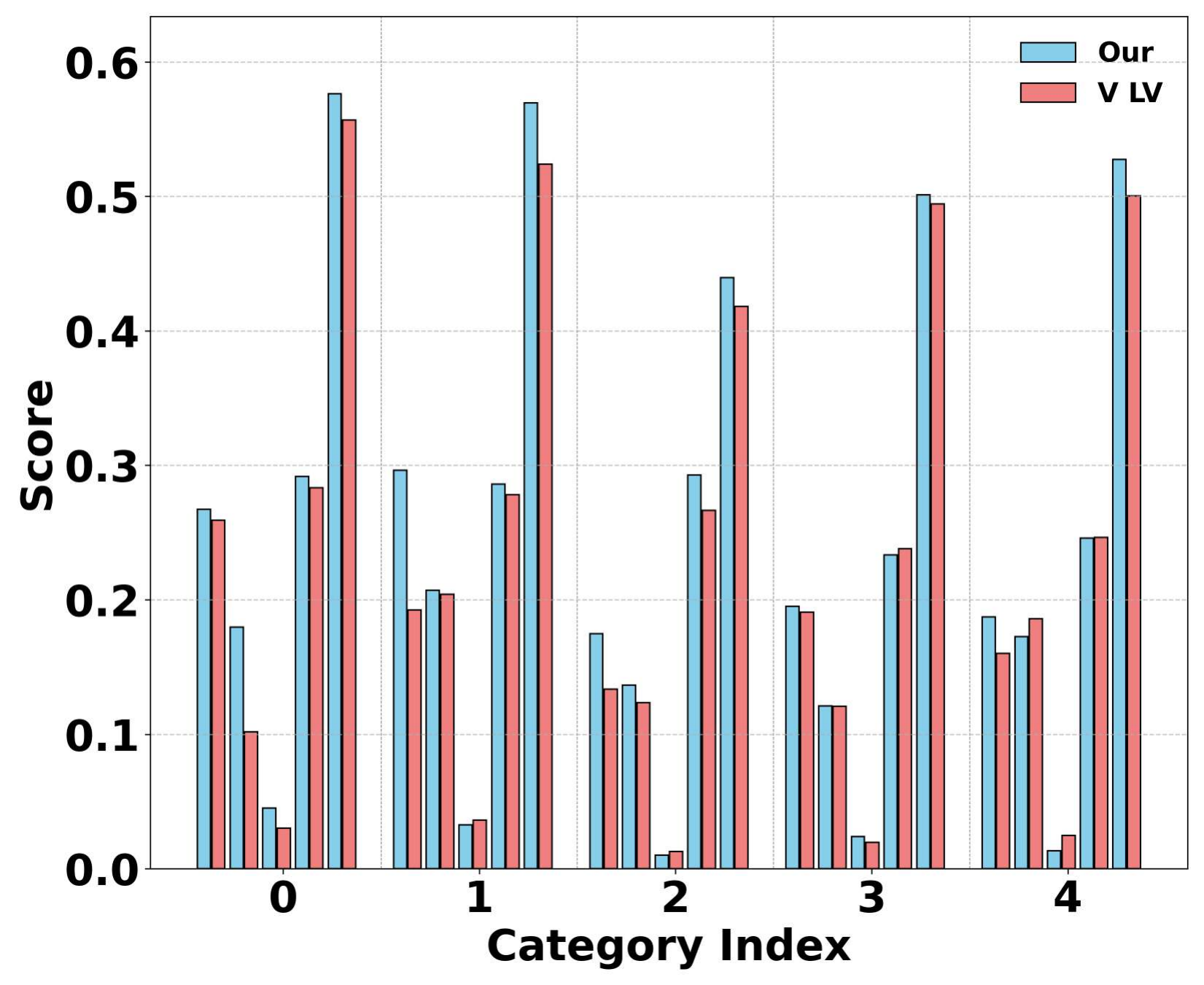}
        \subcaption{Three-person scenario}
        \label{fig:3_person}
    \end{subfigure}

    \vspace{0.5em}

    \begin{subfigure}[b]{0.45\columnwidth}
        \centering
        \includegraphics[width=4.5cm]{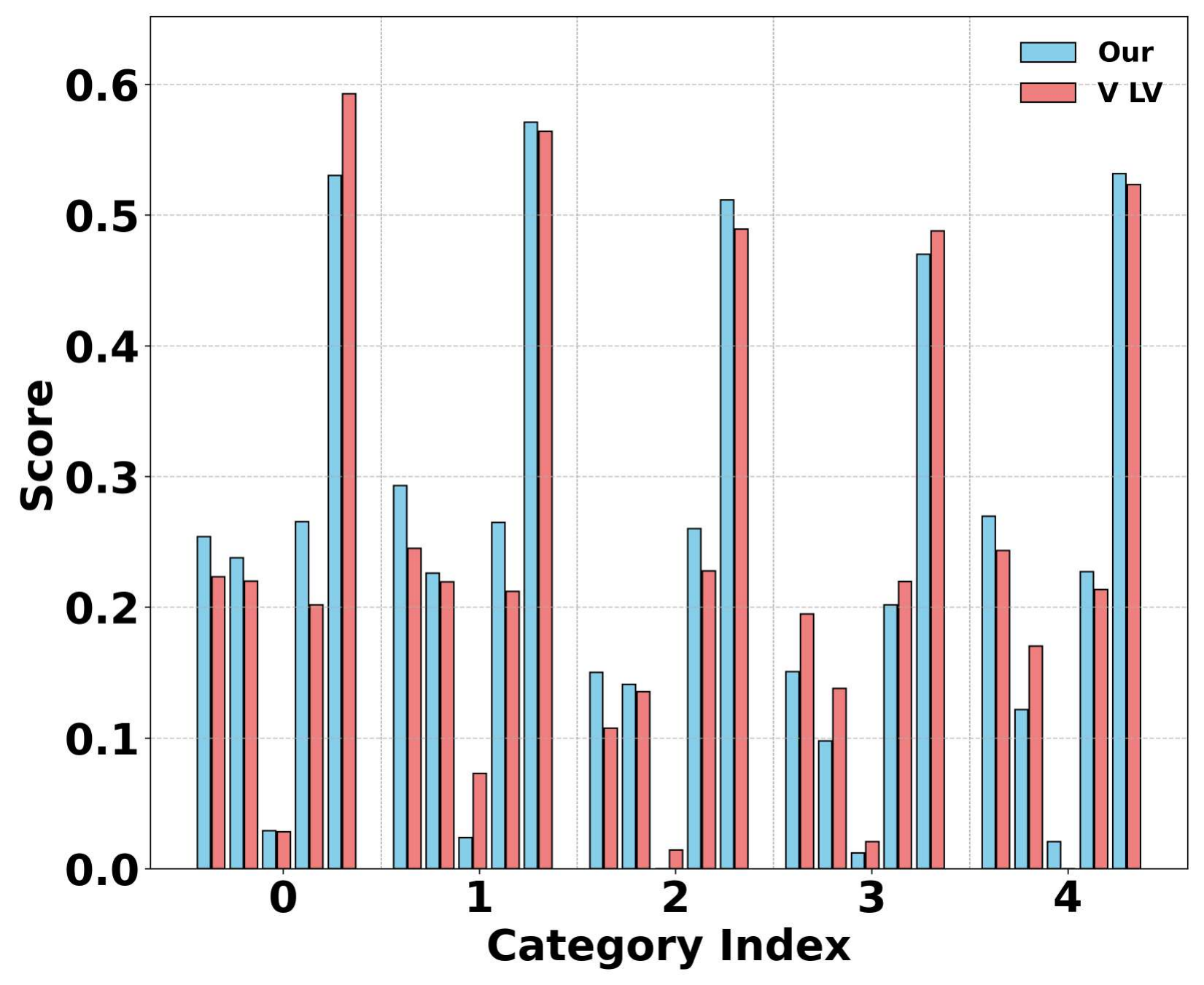}
        \subcaption{Four-person scenario}
        \label{fig:4_person}
    \end{subfigure}

    \caption{Comparison between the proposed scheme and the baseline using automated semantic scoring metrics (ROUGE-1, ROUGE-L, BLEU-4, METEOR, and BERTScore) under multi-person scenarios with increasing scene complexity.}
    \label{fig:automated_score_multi}
\end{figure}

Figures~\ref{fig:2_person}, ~\ref{fig:3_person}, and ~\ref{fig:4_person} represent the automated evaluation results for the two, three, and four-person scenarios, respectively. using the same scoring setting as the single-person scenario (ROGUE-1, ROGUE-L, BLUE, METEOR, and BERTScore). For consistency, these five evaluation metrics are indexed from $0$ to $4$ in all figures. Across all scenarios, our model consistently outperforms the baseline on average across all metrics. While performance naturally degrades as scene complexity increases, these results demonstrate that WirelessSenseLLM maintains stronger robustness and more stable semantic reasoning compared to the baseline. 

\section{Conclusion}\label{AA}
This work presents WirelessSenseLLM, a language-driven wireless sensing scheme that utilizes cross-domain language knowledge to enable zero-shot understanding of human movements directly from CSI measurements. Rather than classifying pre-defined action categories, WirelessSenseLLM generates textual descriptions of body part movements and primitive actions, then uses LLM to infer higher-level activities and answer motion-related queries. By bridging wireless sensing signals and human language, the proposed scheme introduces a new paradigm for wireless human motion understanding. Extensive experiments show that our proposed framework consistently outperforms the baseline under both model-as-judge evaluation and automated scoring metrics.

At the same time, our results reveal important limitations. First, performance degrades as scene complexity increases in multi-person scenarios. Second, interaction reasoning remains challenging even in simpler settings, indicating that fine-grained spatial relationships and interaction cues are hard to recover from CSI signals. In addition, our evaluation is based on a dataset collected under a specific sensing configuration and limited environments, so robustness across different hardware layouts, rooms, and deployment conditions requires further study. Broader baselines for continuous CSI streams and more efficient inference for real-time deployment also remain for future work. We hope this work provides a foundation for future research on stronger interaction reasoning, scalable multi-person understanding, more comprehensive evaluation, and improved computational efficiency. The training code base and annotated dataset will be publicly released.

\bibliographystyle{IEEEtran}
\bibliography{biblography}

\end{document}